# Assessing the usability of the mobile application game, 'Call of Duty'

Charlotte Graves: S4913266
Word Count: 2,510 excluding tables (1,000)

## *(1) INTRODUCTION*

The technological advances in the smartphone application market has expanded at great lengths in recent years (Kaya, Ozturk & Gumussoy 2019). One of the main smartphone applications that often sees record breaking downloads are gaming apps. The gaming application that recently received the most downloads of all time, 100 million in one week, was the mobile game 'Call of Duty' (CoD). With the game generating revenues of over $8 million across app stores, it is important that the users are satisfied with the game and its experience (Cuthbertson, 2019). Unlike other applications, gamers are drawn to gaming applications for their experience more than their functionality. Ensuring that the game is performing with efficiency and of a high level of satisfaction, is key in enhancing the users experience and remaining popular in such a competitive market (Barnett, Harvey and Gatzidis 2018). There are significant advances of mobile applications such as portability and accessibility but the change in architecture has meant that some aspects of design and usability have had to be constrained. Therefore, mobile games need to be more user focused (Nayebi, Desharnais and Abran 2012).

The current study will look to assess the effectiveness of a specific mobile application game. In order to do so, usability techniques will be explored. Moumane, Idri and Abran (2016) suggest four main methods for assessing usability; expert-evaluation, observations, questionnaires and software logging. Due to their research having a strong focus towards mobile applications, these methods will be adopted in this case study. Their research also illustrates the current ISO standards being ISO 9001 and Pie 9.0. These ensure that customers' needs, and requirements are regulated and is a demonstration of continuous improvement. All smartphone applications undergo these regulatory requirements and therefore users should be operating on the highest-level applications. The standard of smartphone software will not be evaluated in this study as it is regarding a universal application game, however it is important to be mindful of differing features and overall performance, (Moumane, Idri & Abran 2016).

THE AIM: To identify any usability issues with the mobile application game 'Call of Duty' and to pose possible solutions as to how these may be resolved.

OBJECTIVES:
1. To make use of a range of evaluative methods to assess the effectiveness of the app from different angles.
2. To make use of a questionnaire to gain both qualitative and qualitative responses on gamers perceptions of the game's effectiveness.
3. To utilise the methods of cognitive walkthrough and think aloud to generate first-hand experiences and subsequent thoughts regarding the mobile game, from the perspective of an expert and a gamer.

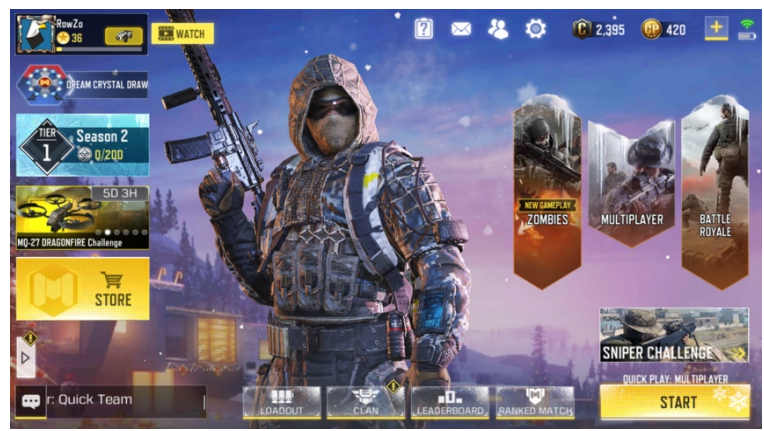

## (2) RICH PICTURE

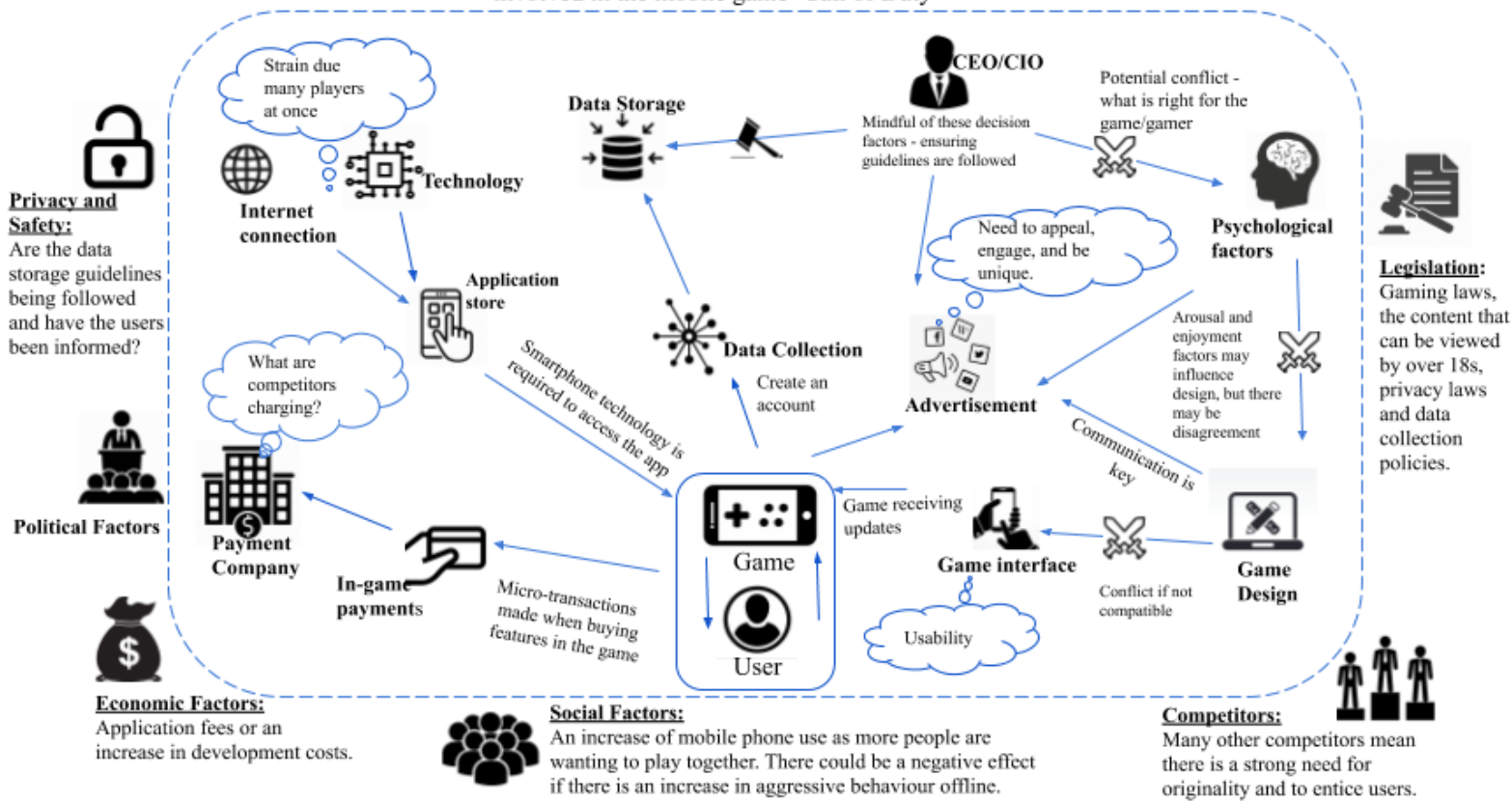

A. A rich picture to display a holistic view of those involved in the running and participation of the mobile application. The picture shows both internal influences such as the game interface and design, as well as in game payments known as 'micro-transactions'. Additionally, there are external players such as payment companies in charge of pricing and payment, as well as data storage systems and subsequent privacy legislation.

## (3) METHODOLOGY

### 3.1) Recruitment

The participants were recruited voluntarily although they were derived from an opportunity sample as they must have knowledge of the game. The participants are required to have an initial level of knowledge as inexperience will affect the results in an inefficient manner. For the samples to be of significant value in terms of this case study, 13 participants completed the questionnaire, 2 experts analysed the cognitive walkthrough to enhance the reliability and 3 gamers underwent the think aloud technique as this is time consuming but enough to uncover rich data. An initial risk assessment and ethics document were signed off as low risk and all participants were asked to provide their consent for the use of their data. All data has been anonymised and followed the Bournemouth university regulations regarding its removable post study.

| Considerations | Nasa TLX | SUS | Cognitive walkthrough | Thinking aloud |
|---|---|---|---|---|
| **Positives** (Holzinger 2005) (Kaikkonen, et al 2005) | Quick and easy method to test workload. Available software to reduce effort. It's a multi-dimensional approach | Versatile to many systems and products. Allows for the collection of Quantitative data – particularly useful for comparative analysis. | Allows for 'experts' to trial the product from a user's perspective. This means that effective identification of problems can be noted. This method can help to gain an understanding of user's perceptions and expectations before seeking their feedback. | Think aloud – verbalise thoughts giving insight into cognitive processes. This method could uncover expectations and misconceptions. Close approximation to the usage. |
| **Negatives** (Holzinger 2005) (Kaikkonen, et al 2005) | Can be intrusive and disruptive. Relies of retrospective data. Response bias, may tailor answers to justify a poor performance. | Subjective measure of one's perception of usability but is often mistaken as objective. It doesn't allow for any elaboration. Any score above 70 is deemed acceptable but it may have issues that are by-passed. | There is a concern of expert bias due to improper task selection. This method is also subjective and therefore must be analysed and compared to an additional experts' response. | This is time consuming and if the user is distracted, they may not initially offer much information. Therefore, they need to be promoted. May be seen as unnatural so performance may be affected. |
| **Decision made** | This method has not been chosen because the case study is not concerned about how the user is dealing with workload or frustration levels. It appears there are more appropriate methods. | This method has not been selected as the case study is not undergoing a comparative analysis and therefore it is deemed more appropriate to conduct a cognitive walkthrough to examine usability in this way. | This method is going to be selected due to the experts being able to gain a sound understanding of the game and its requirements. This will mean that additional methods of analysis can be tailored appropriately. | This method is going to be selected for its ability to gain first-hand, realistic responses from player of the game. For this particular study it is important that the users are engaging, and this reduces retrospective bias. |

*A. Table of possible method considerations and decisions made. Further evaluation can be found in each method section below;*

### 3.2) Questionnaire

Questionnaires are an effective way of gathering a lot of information from a wide sample. The technique is relatively quick and simple to carry out. For the purpose of this study, the questionnaire will make use of both quantitative and qualitative questions to gather a mixture of data, Gupta and Khanna (2017). The qualitative questions will allow for the user's views to be expressed and to gain an insight into their expectations and possible criticisms (Moumane, Idri & Abran 2016). The ability to tailor questions is important for this particular study. General usability questionnaires do not allow for focused questions regarding features of the game. Hussain, Abbas, Abdulwaheed, Mohammed and Abdulhussein (2015) suggests that there are seven factors that affect the usability of smartphone games, four of those being; learnability, memorability, satisfaction, simplicity. These themes will therefore be included in some of the questions that are asked.

### 3.3 Cognitive walkthrough

Desurvire and El-Nasr (2013) explains how heuristics are a good way to measure different aspects of mobile games and therefore make predictions of potential issues that gamers may face. However, this appears more appropriate to developers and designers, therefore in this context, the method of cognitive walkthrough will be used. As explained by Gupta and Khanna (2017) cognitive walkthrough has similar aspects to that of heuristic evaluation as products are analysed against a set of criteria. However, cognitive walkthrough requires an expert to physically complete set tasks while completing an evaluation spreadsheet. This is more appropriate for this study as the expert is able to play the game first-hand and explore the platform in the same way a gamer would. Schoffman, et al (2017) In comparison to heuristics, this method gains feedback from the user rather than applying perceived responses to past standards. This method allows for the effectiveness of the game to be tested. Cognitive walkthrough can be completed by two experts to enhance its credibility. Due to the nature of the game requiring participant involvement, it is important to balance additional methods to compensate and ensure evaluation is not intrusive and gains a variety of feedback responses.

As this case study is assessing a mobile game, the effects of playability need to be addressed, (Desurvire and Wiberg 2009). The authors explain that as well as standard interface design for apps or websites, there are additional concepts to consider with games; for example, immersion, entertainment and challenge. Such principles of playability may appear more obvious to the participant to critique however, it is thought that heuristic evaluation of this topic, would be more beneficial to *designers* during the developmental stage. Therefore, due to the nature of the current task, cognitive walkthrough remains as the chosen method to explore the games features, as the game is past development

### 3.4 Think Aloud

The method of think aloud is often used as a form of analyse when using interactive media. Desurvire and El-Nasr (2013) highlights the importance of thinking aloud in a gaming context. The technique reduces the bias and retrospective nature of self-reports. Thinking aloud is an efficient method to use when analysing initial responses in an immersive context, such as gaming. In some cases, issues may be addressed that would otherwise be overlooked or forgotten at a later stage. As previously mentioned, Moumane, idri and Abran (2016) place *observations* as one of the four main methods used for usability evaluation. In terms of mobile applications, the authors highlight the use of the thinking aloud method to collect rich and representable data. It is suggested that field data collection can lack replicability and control. However, with consideration towards this particular case study, it is important that the user is physically engaging with the game in order to gain a realistic insight into any usability issues.

## (4) RESULTS

### (4.1) Questionnaire Results

*B. A graph to show the percentage of response indicating unnecessary features.*

*C. A graph to show the percentage of responses that felt there were improvement and where they could be made.*

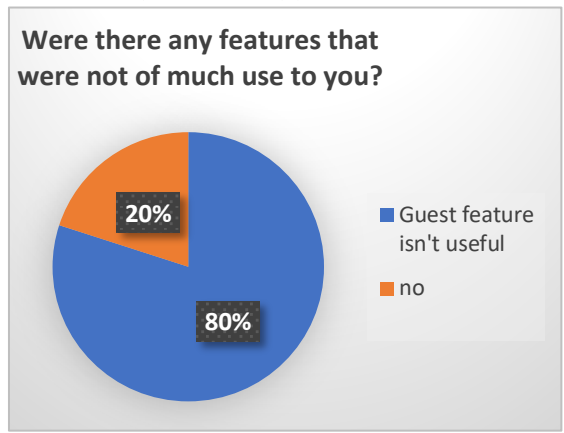

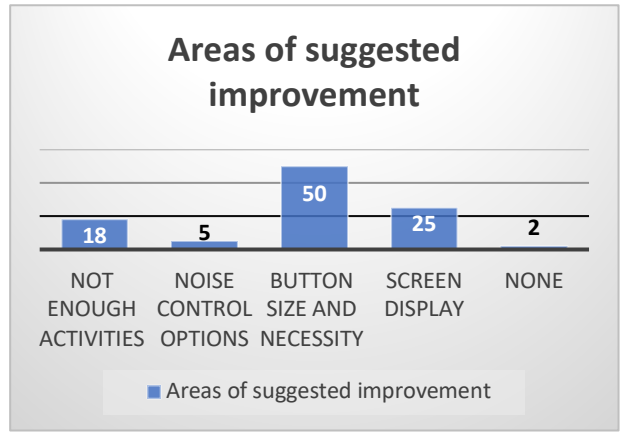

1) The two main findings that appeared most frequently were that of the above. 80% of responses said that they felt there was little need for the guest feature and didn't impact their experience. However, a small number of others didn't report there to be any wasted features.

2) The second graph displays the largest concern that was frequently reported on more than one question; the buttons. The size and location of the buttons have influenced the game play of many of the participants as they are often incorrectly pressed or awkward to adjust. This coincides with the comments regarding the interface being too busy and containing unnecessary features that could be stored elsewhere. Interestingly, two responses reported issues of having physical difficulties regarding the position of their fingers and holding down the buttons.

3) Participants reported that although the sound is important in some conditions, it isn't always necessary and consequently could be distracting. When asked to report possible improvements, a few responses commented on the ability to have a noise control feature which can be modified depending on the condition the player is in.

4) 90% of responses said that the ability to control the difficulty level of the game meant that the game was more enjoyable and rewarding. However, improvements towards an increase in features such as new maps were suggested for enhanced entertainment.

5) 70% of the responses said that they believe the game to be simple enough for non-gamer to pick up, however when asked to elaborate, many reported that there were features 'they just knew', which new players may struggle with in the beginning.

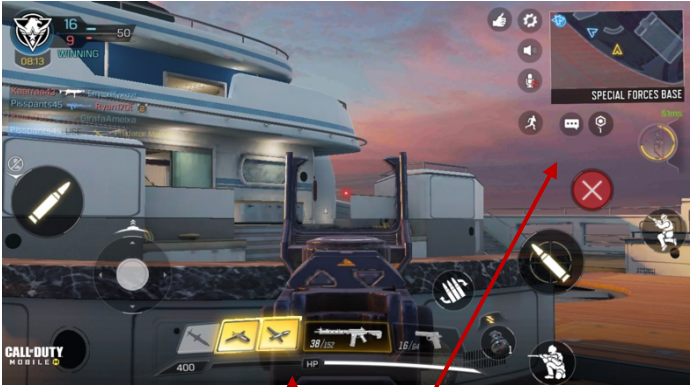 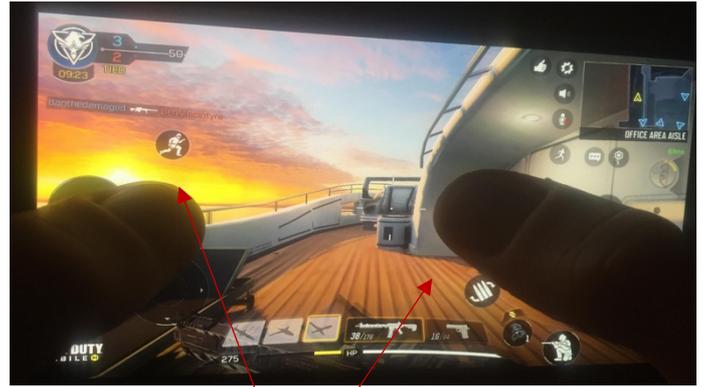

This is the screen in play. The buttons at the top right of the screen are those seen as problematic. Those at the bottom are difficult to press whilst both thumbs are in action.

When the player engages in the game, it can be seen that their view is obstructed from the positioning of their thumbs. The size difference between the thumb and button size can also be seen.

*(4.2) Cognitive Walkthrough Results*

**1|** Will the user know what sub-goal they want to achieve?
**2|** Will the user notice that the correct action is performed?
**3|** Will the user understand that the sub-goal can be achieved by the action?
**4|** Does the user get feedback from the action?

| Tasks | Overall Answers **1\|2\|3\|4** | Feedback from expert 1 | Feedback from expert 2 |
|---|---|---|---|
| **Task 1:** Load into multiplayer and shoot your gun | Y\|Y\|Y\|Y | This task proved self-explanatory and simple to achieve. Each of the sub-categories were easily found and clearly displayed/labelled. The game provided continued feedback through the sub-tasks e.g. noise or colour change confirmation | The task can be completed with ease. Some of the sub-category steps are highlighted on the screen to aid selection. The screen /interface responds with all commands ensuring the player knows each of the buttons have been selected. |
| **Task 2:** Enter a battleground and throw a grenade. | N\|Y\|N\|Y | Issues became apparent in section 1 and 3 of this task. The tools used to execute the tasks were not immediately obvious and were not labelled, this is meant they required practise or prior knowledge. Section 3 saw a lack of understanding; it wasn't clear how the user would achieve the goal and would involve trial and error. | Similar comments were reported. If the user did identify the correct tools, its intention wasn't clear enough meaning the user was left to guess. It was felt that much of the task involved a lot of assumption as opposed to information and any changes that were made, were not made obvious enough. |
| **Task 3:** Adjust the left bullet button**.** | Y\|Y\|N\|Y | The main issue reported with this task was the lack of information available. Similarly, to that of the above, the options that are on the screen are not self- | Prior knowledge is assumed again which may cause confusion for players. It has also been noted that at stage 3, the interface doesn't interact until after the task is |

| | | | explanatory and therefore may cause confusion. The scale is difficult to adjust on the screen. | completed, leaving the player unsure. Similar comments were made about the difficulty of the scale button. |
|---|---|---|---|---|
| **Task 4:** Equip an M4 gun. | | N\|Y\|N\|Y | In the initial stages, there is a lack of labelling or description e.g. there are no clues to suggest which box is the correct one.<br>This task requires specific movements otherwise the commands won't work. This may be frustrating without prior knowledge. | Comments agree that while the game is not being played, more descriptive information could be on the screen. There is too much ambiguity with some of the buttons and the actions that must be performed. |

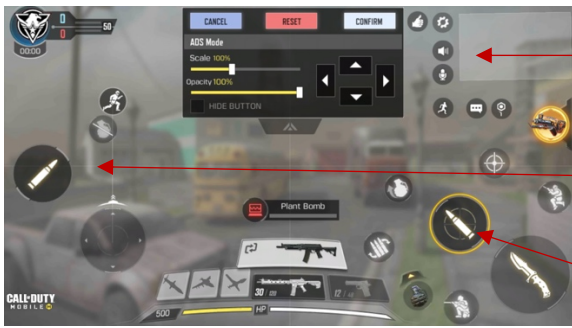

This image helps to illustrate the variety of buttons that players have been reporting as problematic and 'fiddley'. Additionally, it can be seen that are a lack of labels on the majority of the buttons. For new players this may cause difficulties and requires prior knowledge to be efficient. However, availability of this feature has proven essential in allowing player to make adjustment to the buttons. Also, feedback of a selection is evident when highlighted in yellow. This provides confirmation to the player.

*(4.3) Thinking aloud Results*

| Thinking aloud Transcript | Player 1 | Player 2 | Player 3 |
|---|---|---|---|
| | **Expert:** Enter the game when you're ready<br>**Player:** Ok, I'm in the main menu waiting to start a game<br>**Expert:** are there any comments you have at this stage<br>**Player:** no<br>**Expert:** is there anything you wish to comment on<br>**Player:** changing from gun to grenade is sometimes annoying<br>**Expert:** annoying?<br>**Player:** yeah, the buttons are too close and would be annoying to amend while I'm in game.<br>**Player:** It's fun when playing with others I don't really think too much about the buttons until they restrict my view sometimes.<br>**Expert:** what are you doing now?<br>**Gamer:** just deciding which mode to enter<br>**Expert:** does having the option affect the game<br>**Player:** yeah definitely although I'd like more choice, can get boring after a while. | **Expert:** Enter the game when you're ready<br>**Player**: It's taking longer than usual to start as there are a lot of ads to close before i can start.<br>**Player**: targets could be made more obvious, the view is sometimes restricted due to the small screen.<br>**Player:** switching between crouching and lying down is difficult because it's the same button.<br>**Expert**: do you have any comments regarding the sound?<br>**Player:** the sound is fine<br>**Player:** I just found it difficult to change my gun as the button was small and temporary.<br>**Expert**: Do you have any comments regarding navigation? | **Expert:** Enter the game when you're ready<br>**Player:** I've entered a game so I'm just shooting enemies at the moment.<br>**Expert:** Do you have any comment regarding the game play?<br>**Player:** I think its really good quality for a phone game<br>**Expert:** Are there any physical constraints?<br>**Player:** no, I think I've got used to it and don't play the game too long so I don't find it get uncomfortable but some of my friends say it can get straining.<br>**Player:** actually, here's an issue, I've just gone to change my loadout and the button is so small I can barely adjust it.<br>**Expert:** are there any other features similar to this?<br>**Player:** I guess the buttons are small in general but in the main menu you have more time, so it is less frustrating. |

| | **Player:** it can get quite tense and my thumbs start to ache so i don't often play for that long.<br>**Expert:** Do you have any last comments?<br>**Player:** I enjoy the game; it could just be adapted to be less busy on the screen. There are unnecessary features.<br>Player: The game drains my battery a fair bit so it would be good if my battery percentage was shown somewhere so I don't have to exit the game. | **Player**: the navigation is good, but the screen is very full and could be improved. | **Expert:** is there anything else you think you could comment on?<br>**Player:** luckily, I'm familiar with the game but now i think about it the text font is very difficult to read and this can't be altered. This might be an issue for a new player. |
|---|---|---|---|

## *(5) DISCUSSION*

Overall, the game was reported as enjoyable by all participants; most specifically the social and accessibility elements. The questionnaire responses revealed that participants enjoyed the competitive nature of the game and its ability to be tailored to each user. However, there were repeated reports of frustration and suggestions towards areas of improvements. This suggests that the sample were not completely satisfied. After collating the data, all three varying techniques; Questionnaire, cognitive walkthrough and thinking aloud expressed similar findings. Firstly, it was evident that the game is targeted at experienced players. Although the case study accommodated for this by only recruiting those with knowledge of the game, the results highlighted the difficulties that new players might face with usability. In particular the cognitive walkthrough utilised two expert opinions, which took a detailed analysis of the game and Identified the lack of labelling and information to be a possible issue for usability. It was reported that often the sub-tasks resulted in trial and error as movements are based on assumption. This highlights a limitation of the current study. The sample of experienced players was initially justified, as errors due to a lack of knowledge needed to be avoided. However, Holzginer (2005) a comparative study could analyse frustration levels faced by new players compared with experienced ones, with the use of NASA TLX. These findings could identify the exact areas of difficulty for new players rather than placing predictions.

The study did well to collect qualitative data from both a questionnaire and thinking aloud. The questionnaire allowed participants to reflect on the game and consider their answers with more time and without the presence of a researcher. This may mean they were more inclined to provide honest feedback. However, the use of a questionnaire may also mean that participants are responding in line with social desirability and therefore not expressing their true opinions. Similarly, the thinking aloud technique provided the research with findings that were not accounted for elsewhere. The players were engaged in the game and on some occasions, it triggered thoughts that would not have otherwise been expressed. However, the presence of the researcher during the task may have been off-putting and the participants may have held back information due to being nervous or again, social desirability. It would therefore be interesting to conduct these techniques over a larger sample and compare the results. Additionally, a video observation could be conducted as this is less intrusive and overt, therefore natural behaviours may be discovered.
An additional limitation of the study is that it lacks in objective measures and therefore the findings cannot be generalised. Combining this with the small preliminary sample, it is difficult to display these findings with confidence at this stage.

## *(6) CONCLUSIONS*

The mobile application game 'Call of Duty' has received a lot of attention from its record-breaking downloads, this coupled with the increase in smartphone application use, was the rationale behind the case study. The aim of this study was to assess the usability of the application using three techniques. Firstly, usability enquiry; A questionnaire to gather information about the players perceptions and expectations from playing the game and to uncover issues with it being played on a smartphone. Secondly, Usability inspection; A cognitive walkthrough was conducted by two experts to explore areas of the game from the perspective of a gamer in an attempt to identify possible issues. Lastly, Usability Evaluation; Thinking aloud. This was important for this particular case study as it allowed the expert and gamer to interact and observe the game being played live. This method meant that less obvious usability comments were brought to the forefront for later analysis. As Barnett, (2018) suggests that the gaming industry is one of the fasting growing fields, future research could look to compare the usability of the different tools e.g. Mobile, PlayStation, XBOX etc. Interestingly, the accessibility and portability of the game was a significant factor that enticed players to download the app. Therefore, it could be suggested that if more people are admiring such factors, then developers need to be informed of which areas to prioritise during development. In terms of this research, the majority of participants suggests improvements towards the usability and functions of the on-screen buttons. Such information could be vital in maintaining popularity in such a competitive market.